\documentclass[letterpaper, 10 pt, conference]{ieeeconf}
\IEEEoverridecommandlockouts 
\usepackage{graphicx} 
\usepackage{cite}
\usepackage{caption}
\usepackage{amsmath}
\usepackage{amsfonts}
\usepackage{dsfont}
\let\labelindent\relax
\usepackage{enumitem,kantlipsum}
\usepackage{subcaption}
\usepackage{bm}
\usepackage{multirow}
\usepackage{graphicx}
\usepackage{color}
\usepackage[english]{babel}

\usepackage[linesnumbered,ruled,vlined]{algorithm2e}

\date{} 

\title{\LARGE \bf Optimal Sequencing and Motion Control in a Roundabout \\ with Safety Guarantees}

\author{Yingqing Chen, Christos G. Cassandras and Kaiyuan Xu 
\thanks{Y. Chen, C. G. Cassandras and K. Xu are
with the Division of Systems Engineering and Center for Information and
Systems Engineering, Boston University, Brookline, MA 02446
\tt\small\{yqchenn;cgc;xky\}@bu.edu.}
\thanks{This work was supported in part by NSF under grants ECCS-1931600, DMS-1664644, CNS-1645681, and by ARPA-E under grant DE-AR0001282.}
}

\begin{document}
\maketitle
\thispagestyle{empty}
\pagestyle{empty}
\SetKwInput{KwInput}{Input}                
\SetKwInput{KwOutput}{Output} 

\begin{abstract} 
This paper develops a controller for Connected and Automated Vehicles (CAVs) traversing a single-lane roundabout. The controller 
simultaneously determines the optimal sequence and associated optimal motion control jointly minimizing travel time and energy consumption while providing speed-dependent safety guarantees, as well as satisfying velocity and acceleration constraints.
This is achieved by integrating (a) Model Predictive Control (MPC) to enable receding horizon optimization with (b) Control Lyapunov-Barrier Functions (CLBFs) to guarantee convergence to a safe set in finite time, thus providing an extended stability region compared to the use of classic Control Barrier Functions (CBFs).
The proposed MPC-CLBF framework addresses both infeasibility and myopic control issues commonly encountered when controlling CAVs over multiple interconnected control zones in a traffic network, which has been a limitation of prior work on CAVs going through roundabouts, while still providing safety guarantees.
Simulations under varying traffic demands demonstrate the controller's effectiveness and stability.
\end{abstract}




\section{Introduction}
It has been well documented that urban congestion has reached critical levels over the past few decades in terms of time, pollution, and fuel consumption \cite{chang2017there}. 
The primary bottlenecks that contribute to such congestion are conflict areas such as intersections, roundabouts, merging roadways, and speed reduction zones\cite{rios2016automated}.
The emergence of Connected and Automated Vehicles (CAVs) along with real-time communication between mobile endpoints and the infrastructure \cite{liu2023} make it possible to achieve  smoother traffic flow and lower fuel consumption \cite{campi2023roundabouts} through better information utilization and more precise trajectory design.

Most research to date has focused on the control and coordination of CAVs within a single \emph{Control Zone} (CZ) that encompasses a conflict area such as merging roadways \cite{milanes2010automated},\cite{xiao2021decentralized}, lane changing \cite{armijos2022sequential}, \cite{li2023cooperative} and unsignalized intersections \cite{bichiou2018developing},\cite{zhang2019decentralized}. However, the transition from a single CZ to multiple interconnected CZs is particularly challenging \cite{xu_scaling_2023}: while in an isolated CZ it is assumed that vehicle states initially satisfy all constraints upon entering this CZ, 
when several CZs are interconnected it is generally the case that the state of a vehicle exiting one CZ does not satisfy the next CZ's constraints. Additionally, since the traffic flow is propagated throughout a network of CZs, myopic optimal control limited to one CZ may require extra control effort in the next
and even cause congestion if CZs are in close proximity, resulting in blocked traffic. Thus, directly applying control techniques used in a single CZ to multiple CZs without considering system-wide or local neighboring traffic information
leads to performance degradation and lack of safety guarantees,
especially when the CZs are tightly coupled.

A typical case in point of interconnected CZs arises in a roundabout, a configuration which is attractive to traffic control because its geometry enhances safety by promoting
lower speeds, better visibility, increased reaction time for
drivers, and resulting in less severe crashes when they do occur \cite{rodegerdts2010roundabouts}. Though roundabouts offer many such benefits, their effective
implementation requires more careful planning, since the compact geometry with multiple closely-spaced Merging Points (MPs) 
requires tighter control and more stringent safety constraint satisfaction. Both model-based and learning-based methods have been considered to deal with the control and coordination of CAVs at roundabouts.
The latter try to learn effective trajectories along the entire route; for example,
\cite{garcia_cuenca_autonomous_2019} proposes a Q-learning framework for sequential and automatic decision-making of autonomous vehicles driving in a roundabout. The reward is set to a deviation angle so that lane tracking is learned
regardless of other efficiency metrics. The cooperation between
CAVs is considered in \cite{capasso_intelligent_2020} where the model is trained in a cooperative multi-agent fashion in order to execute a maneuver interpreting the intention of other drivers.
On the other hand, model-based mechanisms often apply the optimal control framework to optimize the trajectory through the MPs of roundabouts while avoiding potential collisions. 
Some works break the problem down into several cruise control periods with fixed terminal time determined by rules \cite{farkas_mpc_2022},\cite{martin-gasulla_single-lane_2019} or an optimization model \cite{bakibillah_bi-level_2021},\cite{bichiou_developing_2019}, ignoring the coupled CZ structure.
Other works consider the entire trip within the roundabout as a complete optimal control problem.
For example, \cite{zhao_optimal_2018} proposes a decentralized optimal controller for CAVs in a roundabout which minimizes the control input while satisfying rear-end safety constraints assuming a First-In-First-Out (FIFO) queue in the entire control zone. In \cite{chalaki_experimental_2020} the multi-lane roundabout problem is decomposed to optimize both energy and traveling time assuming no constraint is violated. In order to improve computational efficiency, a joint Optimal Control and Control Barrier Function (OCBF) approach is used in \cite{xu2021decentralized} such that unconstrained control trajectories are optimally tracked while also guaranteeing the satisfaction of all constraints using Control Barrier Functions (CBFs). In addition, in order to optimize the estimated future performance, Model Predictive Control (MPC) is used in \cite{nor2018merging} and \cite{farkas_mpc_2022}. 

However, these works fall short of addressing the control issues that arise in interconnected CZs as described earlier.  Learning-based methods are reward-driven and cannot reliably prevent conflicts between vehicles. Among model-based techniques, some fail to account for the tightly coupled conflict structure and regard a roundabout as consisting of multiple independent CZs, prone to safety violation, energy waste, and congestion. 
Our previous work \cite{xu_scaling_2023} deals with the infeasibility caused by constraint violations as vehicles progress through successive CZs by defining a Feasibility Enforcement (FE) mode which enforces the satisfaction of initial constraints as fast as possible within a CZ by setting a maximum deceleration. This approach sacrifices considerable energy to enforce feasibility, an inefficiency that can be avoided with proper control. Therefore, the first contribution of this paper is to propose a decentralized MPC-CLBF framework which leverages the coupled structure of multiple interconnected CZs and addresses both infeasibility and myopic control issues. In particular,
building on the use of optimal control and CBF-based methods in \cite{xiao2021decentralized}, we formulate an optimal control problem for CAVs traveling through a roundabout where constraints are formulated using \emph{Control Lyapunov-Barrier Functions} (CLBFs) \cite{xiao2021high}. When
constraints from the next CZ are not satisfied for vehicles entering that CZ, CLBFs are used to ensure convergence back to a safe set in \emph{finite} time, thus providing an extended stability region relative to classic CBFs. In order to overcome the myopic nature of prior CBF-based controllers, Model Predictive Control (MPC) is used to account for future performance across different CZs, therefore achieving optimality over a tunable receding horizon.

The second contribution of this paper is to consider the joint solution of the \emph{sequencing} and motion control problems. The problem of sequencing vehicles through a roundabout has been dealt with by mostly assuming that CAVs maintain a FIFO order, which has been shown to be often inefficient in terms of both travel time and energy consumption \cite{xu2021comparison}. This assumption is relaxed in \cite{xu2021decentralized} and replaced by a Shortest-Distance-First (SDF) sequence where the distance to the next MP as well as the current vehicle speed are considered. In addition to fixed rules, Dynamic Resequencing (DR)\cite{zhang_decentralized_2018} and Optimal Dynamic Resequencing \cite{xiao_decentralized_2020} have been proposed by selecting the sequence maximizing performance over a single merging maneuver. While dynamic resequencing methods have shown improvement over fixed sequences, local optimal sequences fail to consider control after merging, thus they are not able to ensure optimal coordination over interconnected CZs. In our proposed solution of the joint sequencing and motion control problems we combine an optimal sequencing process with the MPC-CLBF framework mentioned above so as to select an optimal sequence and motion control of each CAV in this sequence to jointly optimize the future vehicle speed and its energy consumption while guaranteeing safety. 

The paper is organized as follows. In Section \ref{Sec:Problem Formulation}, the roundabout problem is formulated as an optimal control problem with safety constraints. In Section \ref{Sec:Optimal Resequencing}, an optimal sequencing approach using the proposed MPC-CLBF framework is developed. Simulation results are presented in Section \ref{Sec: Simulation} showing effectiveness of our controller. Finally, in Section \ref{sec: conclusion} we provide conclusions and future research.

\section{Problem Formulation}\label{Sec:Problem Formulation}
\begin{figure}
    \centering
    \includegraphics[width=0.7\columnwidth]{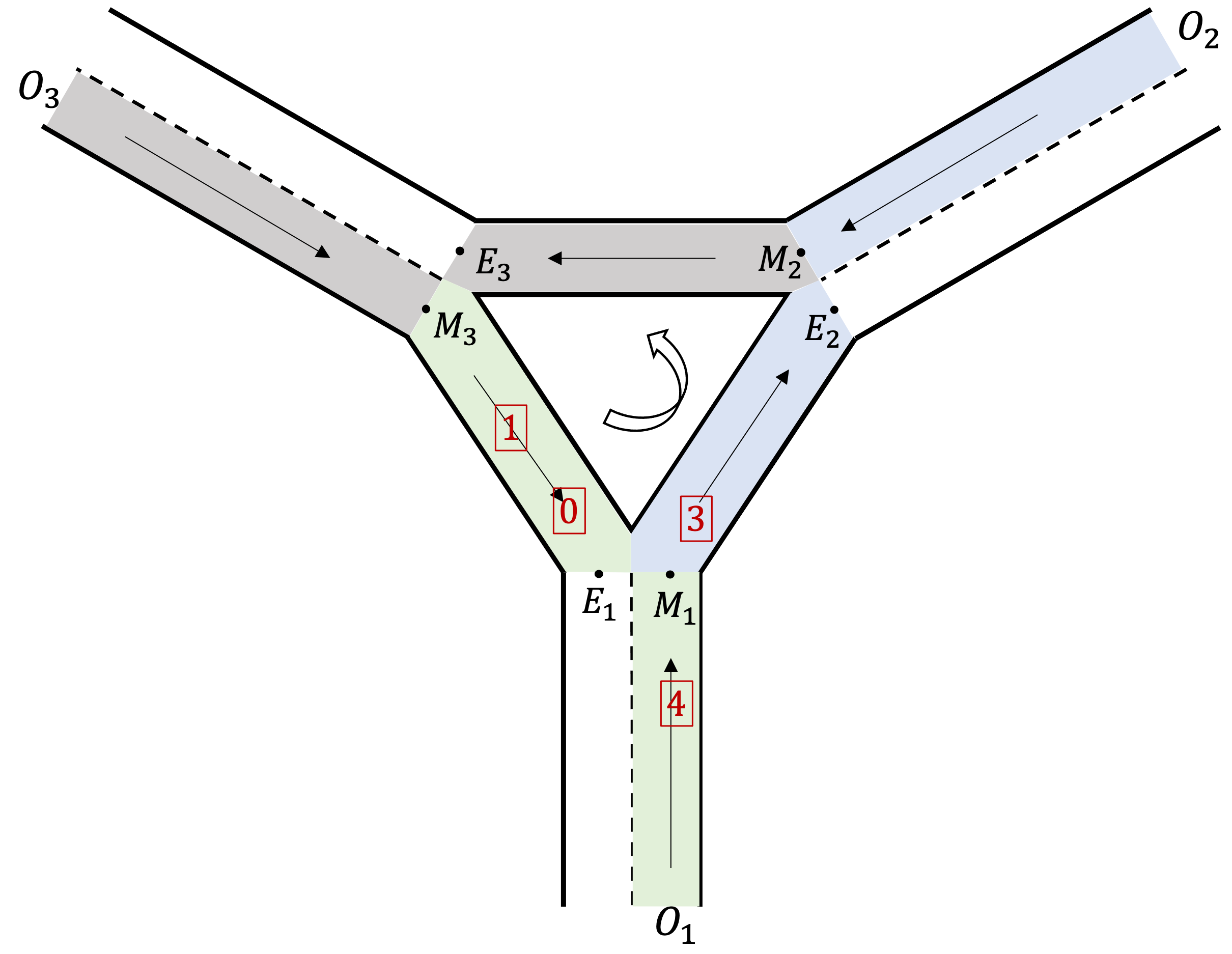}
    \caption{A roundabout with 3 entries }
    \label{fig:roundabout}
\end{figure}

We consider a single-lane triangle-shaped roundabout with $N$ entry and $N$ exit points where $N\geq 2$. For simplicity we limit our analysis here to $N=3$, as depicted in Fig. \ref{fig:roundabout}, with straightforward extensions to $N>3$.
We chose a triangular geometry to simplify the analysis, focusing on the joint sequencing and motion control problem; the extension to circular roundabouts complicates things, but can still be handled as 
demonstrated in \cite{xu2022decentralized}. Additionally, implementing a triangular layout enables us to carry out direct performance comparisons to prior results using a similar configuration in \cite{xu2021decentralized} so as to evaluate the advantages of the joint sequencing and motion control approach developed in this paper with all traffic consisting of CAVs. Thus, as in \cite{xu2021decentralized},
we consider the case where all traffic consists of CAVs which randomly enter the roundabout from three different origins $O_1$, $O_2$ and $O_3$ and have randomly assigned exit points $E_1$, $E_2$ and $E_3$. We assume all CAVs move in a counterclockwise way in the roundabout. 

The entry road segments are connected with the triangle at the three Merging Points (MPs) labeled as $M_1$, $M_2$ and $M_3$, where CAVs from different road segments may potentially collide with each other. Centered at each MP, we partition the roundabout into three Control Zones (CZs), labeled $CZ_1$, $CZ_2$ and $CZ_3$, and shaded in green, blue, grey respectively in Fig.\ref{fig:roundabout}. Each CZ includes two road segments towards the corresponding MP: one is the entry road indexed by $1$, and the other is the segment within the roundabout, indexed by $0$. Thus, we define $c_i \in \{0,1\}$ as the road segment where CAV $i$ is currently located, where $c_i=1$ indicates CAV $i$ is at an entry road, otherwise $c_i=0$.
We also assume that all road segments have the same length $L$ (extensions to different lengths are straightforward). The full trajectory of a CAV in terms of the MPs it must go through can be determined by its entry and exit points.

The vehicle dynamics for each CAV $i\in S(t)$ along the road segment to which it belongs take the form
\begin{equation}
\left[
\begin{array}
[c]{c}%
\dot{x}_{i}(t)\\
\dot{v}_{i}(t)
\end{array}
\right]  =\left[
\begin{array}
[c]{c}%
v_{i}(t)\\
u_{i}(t)
\end{array}
\right]  \label{VehicleDynamics}%
\end{equation}
where $x_{i}(t)$ denotes the distance from the origin of the road segment where CAV $i$ currently locates, $v_{i}(t)$ denotes the velocity, and $u_{i}(t)$ denotes the control input (acceleration).

Let $S(t)$ be the set of indices of CAVs present in the roundabout at time $t$. The cardinality of $S(t)$ is denoted by $N(t)$. When a new CAV arrives at the roundabout, it is assigned the index $N(t) + 1$. Each time a CAV $i$ leaves the roundabout, it is removed from $S(t)$ and all CAV indices larger than $i$ decrease by $1$. We then partition $S(t)$ into subsets based on their current CZ and define a merging group as a set of CAVs present at the same CZ at time $t$. Let $S_k(t)$, $k \in \{1,2,3\}$, be the set of indices of CAVs (from $S(t)$) within the merging group of $CZ_k$ at time t, where $S_1(t)\cap S_2(t) \cap S_3(t) = \emptyset$, $S(t)=S_1(t)\cup S_2(t) \cup S_3(t)$. The cardinality of $S_k(t)$ is denoted by $N_k(t)$ with $k\in \{1,2,3\}$.

\textbf{The Coordinator Table}. A coordinator table is used for each CZ to record the essential state information for all CAVs within it, and to identify all conflicting CAVs under a given merging sequence. An example of such table corresponding to Fig. \ref{fig:roundabout} for $CZ_1$ is shown in Table \ref{table:queue table S_1}. The definition of each column is as follows:
\emph{idx} is the unique CAV index at the roundabout;
\emph{state} is the CAV state $\mathbf{x_i} = (x_i, v_i)$ where $x_i$ is the distance to the location of CAV $i$ from the entry point of its current road segment and $v_i$ is the velocity of CAV $i$;
\emph{initial CZ} is the index of the CZ from which CAV $i$ enters the roundabout;
\emph{final CZ} is the index of the CZ through which CAV $i$ exits the roundabout;
\emph{current CZ} is the index of the CZ where CAV $i$ is currently located;
\emph{$c_i$} is the classification index of the road segement where CAV $i$ is currently located: $c_i=1$ for an entry segment and $c_i=0$ for the segment within the roundabout;
\emph{$i_p$} is the index of the CAV which physically immediately precedes CAV $i$ in the roundabout (if one exists), noting that CAV $i_p$ is not necessarily located in the same CZ as CAV $i$;
\emph{$i_m$} is the index of the last CAV that precedes CAV $i$ (if one exists) at the next MP within a given sequence. CAV $i_m$ must be in the same CZ but a different road segment than that of CAV $i$, i.e., $c_i \ne c_{i_m}$.


\begin{table}[]
\centering
\caption{The Coordinator Table $S_1(t)$}
\resizebox{0.5\textwidth}{!}{%
\begin{tabular}{|llllllll|}
\hline
\multicolumn{8}{|l|}{$S_1(t)$} \\ \hline
\multicolumn{1}{|l|}{idx} &
  \multicolumn{1}{l|}{state} &
  \multicolumn{1}{l|}{initial CZ} &
  \multicolumn{1}{l|}{final CZ} &
  \multicolumn{1}{l|}{current CZ} &
  \multicolumn{1}{l|}{$c_i$} &
  \multicolumn{1}{l|}{$i_p$} &
  $i_m$ \\ \hline
\multicolumn{1}{|l|}{0} &
  \multicolumn{1}{l|}{\pmb{$x_0$}} &
  \multicolumn{1}{l|}{3} &
  \multicolumn{1}{l|}{1} &
  \multicolumn{1}{l|}{1} &
  \multicolumn{1}{l|}{0} &
  \multicolumn{1}{l|}{} &
   \\ \hline
\multicolumn{1}{|l|}{1} &
  \multicolumn{1}{l|}{\pmb{$x_1$}} &
  \multicolumn{1}{l|}{3} &
  \multicolumn{1}{l|}{2} &
  \multicolumn{1}{l|}{1} &
  \multicolumn{1}{l|}{0} &
  \multicolumn{1}{l|}{0} &
  4 \\ \hline
\multicolumn{1}{|l|}{4} &
  \multicolumn{1}{l|}{\pmb{$x_4$}} &
  \multicolumn{1}{l|}{1} &
  \multicolumn{1}{l|}{2} &
  \multicolumn{1}{l|}{1} &
  \multicolumn{1}{l|}{1} &
  \multicolumn{1}{l|}{3} &
   \\ \hline
\end{tabular}%
}
\label{table:queue table S_1}
\end{table}

\textbf{Event-driven Update Process for $S_k(t)$}: 
The three coordinator tables $S_k(t)$, $k\in\{1,2,3\},$ are updated simultaneously in an event-driven manner. The triggering events, and their corresponding update processes are as follows: 
\begin{itemize}
    \item A new CAV enters the roundabout: The CAV is indexed and added to the bottom of the corresponding table. 
    \item CAV $i$ exits the roundabout: All information related to CAV $i$ is removed from the corresponding table. All rows from all tables with index larger than $i$ decrease their index values by 1. 
    \item CAV $i$ exits $CZ_j$ and enters $CZ_{j^+}$, where $j\in \{1,2,3\}$ and $j^+ \equiv j+1 \pmod{3}$: All information of CAV $i$ is removed from $S_j(t)$, and the latest information of CAV $i$ is added to the $S_{j^+}(t)$. 
\end{itemize}

In addition to row insertion and deletion, the columns are updated accordingly, only after any one of the events above occurs. In practice, this process may also be triggered upon a fixed timeout to ensure that no event has been missed due to random disturbances. 
With the information from the coordinator table, the motion control of each CAV can be determined by solving an optimal control problem. 

\subsection{Optimal Control Problem}
We consider two objectives for each CAV subject to four constraints, as detailed next.

$\textbf{Objective 1}$ Minimize the travel time $J_{i,1} = t^f_i - t_i^0$ where $t_i^0$ and $t^f_i$ are the times CAV $i$ enters and exits the roundabout.

$\textbf{Objective 2}$ Minimizing energy consumption:
\begin{equation}
\label{energycost}
\setlength{\abovedisplayskip}{2pt}\setlength{\belowdisplayskip}{2pt}
J_{i,2}=\int_{t_{i}^{0}}^{t_{i}^{f}}C(u_{i}(t))dt,
\end{equation}
where $C(\cdot)$ is a strictly increasing function of its argument. Since the energy consumption rate is a monotonic function of the acceleration, we adopt this general form to achieve energy efficiency.

$\textbf{Constraint 1}$ (Rear end safety constraints): Let
$i_{p}$ denote the index of the CAV which physically immediately precedes $i$ in the roundabout (if one is present), whether they are at the same CZ or not. We define the distance $z_{i,i_{p}}(t):=\bar x_{i_{p}}(t)-x_{i}(t)$ where
$\bar x_{i_{p}}(t)=x_{i_{p}}(t) + L \delta$ if CAV $i_p$ and CAV $i$ are at different CZs and the CZ difference is $\delta$ (taken counterclockwise); otherwise $\bar x_{i_{p}}(t)=x_{i_{p}}(t)$. We require that
\begin{equation}
\setlength{\abovedisplayskip}{2pt}\setlength{\belowdisplayskip}{2pt}
z_{i,i_{p}}(t)\geq\varphi v_{i}(t)+\delta,\text{ \ }\forall t\in\lbrack
t_{i}^{0},t_{i}^{f}], \label{equ: rear-end safety const}
\end{equation}
where 
$v_{i}(t)$ is the speed of CAV $i\in S(t)$ and $\varphi$ denotes the reaction time (as a rule, $\varphi=1.8$s is used, e.g., \cite{Vogel2003}). If we define $z_{i,i_{p}}$ to be the distance from
the center of CAV $i$ to the center of CAV $i_{p}$, then $\delta$ is a
constant determined by the length of these two CAVs (generally dependent on
$i$ and $i_{p}$ but taken to be a constant for simplicity).

$\textbf{Constraint 2}$ (Safe merging constraint):
Let $i_m$ denote the index of the CAV traveling on a different road segment of the same CZ that shares the same next MP, $M_k$, as CAV $i$. CAV $i_m$ directly precedes CAV $i$ in arriving at $M_k$ among all vehicles in its segment under a given crossing sequence. Let $t_{i_m}^k, k \in\{1, 2, 3\}$ be the arrival time of CAV $i_m$ at MP $M_k$.
The distance between $i_m$ and $i$, given by $z_{i,i_m} (t)\equiv x_{i_m} (t) - x_i(t)$, is constrained by
\begin{equation}
z_{i,i_m}(t_{i_m}^{k})\geq\varphi v_{i}(t_{i_m}^{k
})+\delta,  \forall i\in S(t), k\in \{1,2,3\} \label{equ: Merging safe}%
\end{equation}

$\textbf{Constraint 3}$ (Vehicle limitations): There are constraints
on the speed and acceleration for each $i\in S(t)$, i.e.,
\begin{align}
v_{\min} \leq v_i(t)\leq v_{\max}, \forall t\in[t_i^0,t_i^f],\\
u_{i,\min}\leq u_i(t)\leq u_{i,\max}, \forall t\in[t_i^0,t_i^f],  \label{equ:VehicleConstraints}%
\end{align}
where $v_{\max}>0$ and $v_{\min}\geq0$ denote the maximum and minimum speed
allowed in the roundabout, while $u_{i,\min}<0$ and $u_{i,\max}>0$ denote the minimum
and maximum control input, respectively.

\textbf{Problem 1: } Our goal is to determine a control law to achieve
objectives 1-2 subject to constraints 1-3 for each $i\in S(t)$ governed by the
dynamics (\ref{VehicleDynamics}). 
Thus, we combine Objectives 1-2 by constructing a convex combination
as follows: {
\begin{equation}\small
\begin{aligned}J_i=\int_{t_i^0}^{t_i^f}\left(\alpha + \frac{(1-\alpha)\frac{1}{2}u_i^2(t)}{\frac{1}{2}\max \{u_{i,\max}^2, u_{i,\min}^2\}}\right)dt \end{aligned}.\label{eqn:energyobja}%
\end{equation}
}where we set $C(u_{i}(t))=(1/2)u_{i}^{2}(t)$ in (\ref{energycost}) and use the weight $\alpha\in\lbrack0,1]$.
If $\alpha=1$, then we solve (\ref{eqn:energyobja}) as a minimum time
problem. Otherwise, by defining $\beta:=\frac{\alpha\max\{u_{i,\max}%
^{2},u_{i,\min}^{2}\}}{2(1-\alpha)}$ and multiplying (\ref{eqn:energyobja}) by
$\frac{\beta}{\alpha}$, we have: {\small
\begin{equation}
\setlength{\abovedisplayskip}{1pt}\setlength{\belowdisplayskip}{1pt}J_{i}%
(u_{i}(t)):=\beta(t_{i}^{f}-t_{i}^{0})+\int_{t_{i}^{0}}^{t_{i}^{f}%
}\frac{1}{2}u_{i}^{2}(t)dt,\label{eqn:energyobj}%
\end{equation}
}where $\beta\geq0$ is a weight factor that can be adjusted to penalize
travel time relative to the energy cost, {subject to
(\ref{VehicleDynamics}), (\ref{equ: rear-end safety const})-(\ref{equ:VehicleConstraints}) given $t_{i}^{0},x_{i}(t_{i}%
^{0}),v_{i}(t_{i}^{0})$.}

Obviously, solving this problem for any CAV $i$ requires knowledge of $i_p$ and $i_m$ in (\ref{equ: rear-end safety const})-(\ref{equ: Merging safe}) whose assignment is dynamically changing, especially when a CAV changes CZs. Our goal is to determine an optimal sequence for each merging group of CAVs in a CZ associated to a MP while also determining the corresponding optimal control $u_i(t)$.
This requires the presence of a 
coordinator (normally a Road Side Unit (RSU) associated with
the roundabout) which records the information associated with each CAV as described earlier in this section,  updates it as needed and is responsible for determining the optimal merging sequence for each merging group in a CZ. 
We note that while the merging sequence determination is centralized at this coordinator, the evaluation of each CAV's optimal trajectory under a specific merging sequence is performed in a decentralized manner as detailed next.

\section{Optimal Sequencing and Motion Control}\label{Sec:Optimal Resequencing}
Our objective is to assign each CAV an appropriate order within its merging group with respect to the next MP and associated motion control so as to minimize the total travelling time and energy while satisfying the safety constraints (\ref{equ: rear-end safety const})-(\ref{equ: Merging safe}). To solve this problem, we propose a joint process of optimal sequencing and motion control that comprises three parts: (a) determine all feasible sequences that all CAVs in the merging group can follow when crossing the MP, (b) evaluate each candidate feasible sequence using the MPC-CLBF framework, and (c) determine the optimal sequence and associated control for each CAV.  These three steps are detailed next.

\textbf{Feasible Sequences}. A feasible sequence is a prioritized list of indices for all CAVs currently within the same CZ, ordered by projected CZ exit times (either progressing to the next CZ or leaving the roundabout). 
Given the single lane assumption (which excludes overtaking), the sequence must maintain relative on-road precedence. The CAVs not located at their final CZ can conflict with CAVs from other inbound road segments heading to the shared next MP. Thus, the relative merging order of CAVs is flexible, comprising the feasible set of sequences to evaluate. We denote the set of all feasible sequences for $CZ_k$ at time $t$ as $F_k(t)$.
For example, the feasible sequences for $S_1(t)$ as shown in Table \ref{table:queue table S_1} are $F_1(t) = \{[0,1,4], [0,4,1],[4,0,1]\}$. Each feasible sequence is associated with a pair of $i_p$, $i_m$ assignments.

\textbf{Assign $i_p, i_m$.} For each feasible sequence $\bold f\in F_k(t)$,
we first partition it into two subsequences, $\bold f_0$ and $\bold f_1$, based on the $c_i$ value of each CAV, while preserving the original order: $i \in \bold f$ is added to $\bold f_0$ if $c_i=0$, otherwise it is added to $\bold f_1$.
For example, in Table \ref{table:queue table S_1}, $\bold f=[0,4,1]$ is partitioned into $\bold f_0=[0,1]$ and $\bold f_1=[4]$. Then, $i_m$ is assigned to the closest index that is previous to $i$ in $\bold f$ while belonging to the different subsequence than $i$, e.g., $1_m=4$.

Regarding $i_p$, there are two cases.
First, if $i$ does not rank first in its subsequence, its $i_p$ is the index immediately prior to $i$ in this subsequence (e.g., $1_p=0$). Otherwise, 
i.e., it ranks first, then there are two cases to consider:
(a) If $i$'s current CZ is its final CZ (indicating that it is leaving the roundabout), then $i_p$ does not exist (e.g., $0_p$ does not exist).
(b) Otherwise, we need to determine $i_p$ by finding a CAV located in the next CZ which precedes CAV $i$ immediately. If the current CZ is $k$, the next one is denoted by $k^+ \equiv k+1 \pmod{3}$. If there exists some $j \in S_{k^+}(t)$ such that $c_j=0$ and $j$ does not exist in the $i_p$ column of $S_{k^+}(t)$ (which indicates CAV $j$ is the last one on its road segment), 
we set $i_p = j$ (e.g., $4_p=3$ in Fig. \ref{fig:roundabout}). If not, we move to the next CZ and check CAV $i$ following the same logic until we return to the initial CZ, indicating that no $i_p$ exists. 

\textbf{Evaluate a Feasible Sequence}. For each merging group in $CZ_k, k\in\{1,2,3\}$, we need to evaluate each feasible sequence $\bold f \in F_k(t)$ determined above, which is associated with a specific pair of $i_p$ and $i_m$ assignments for every $i$ in this sequence. Thus, we need to evaluate the performance of each CAV $i \in \bold f$ under optimal motion control.

\subsubsection{Optimal Control of Single CAV} \label{subsubsec:Procedure for Evaluating Single CAV} 
We start with the \emph{unconstrained} optimal control solution of 
(\ref{eqn:energyobj}) when safety constraints \eqref{equ: rear-end safety const} and \eqref{equ: Merging safe} are inactive. This has been derived in prior work \cite{xu2021decentralized} based on a standard Hamiltonian analysis \cite{bryson2018applied}, yielding the optimal control, velocity, and position trajectories of CAV $i$:
\begin{align}
    \label{equ:uc_u}
    u_i(t) &= a_i t + b_i \\
    \label{equ:uc_v}
    v_i(t) &= 1/2 \cdot a_i t^2 + b_i t + c_i \\
    \label{equ:uc_x}
    x_i(t) &= 1/6 \cdot a_i t^3 + 1/2 \cdot b_i t^2 + c_i t + d_i
\end{align}
where the parameters $a_i, b_i, c_i, d_i$ and $t_i^f$ are obtained by solving the following five algebraic equations: 
\begin{align}
    \label{eqn: ucp_1}
    1/2 \cdot a_i (t_i^0)^2 + b_i t_i^0 + c_i &= v_i^0\\
    \label{eqn: ucp_2}
    1/6 a_i \cdot(t_i^0)^3 + 1/2 b_i (t_i^0)^2 + c_i t_i^0 + d_i &= 0 \\
    \label{eqn: ucp_3}
    1/6 a_i \cdot(t_i^f)^3 + 1/2 b_i (t_i^f)^2 + c_i t_i^f + d_i &= x_i(t_i^f) \\
    \label{eqn: ucp_4}
    a_it_i^f + b_i &= 0\\
    \label{eqn: ucp_5}
    \beta + 1/2a_i^2\cdot(t_i^f)^2 + a_ib_it_i^f + a_ic_i &= 0
\end{align}
which can be solved efficiently. Therefore, when CAV $i$ has no $i_p$ and no $i_m$, the solution of \eqref{equ:uc_u}-\eqref{eqn: ucp_5} provides the optimal control and state trajectories for evaluation. 

However, if $i_p$ or $i_m$ exists, the constrained optimal control problem becomes hard to solve as multiple constraints can become active. To overcome this problem, Control Barrier Functions (CBFs) have been used 
\cite{xiao2023safe} for three main reasons: (a) The original constraints (\ref{equ: rear-end safety const})-(\ref{equ:VehicleConstraints}) can be replaced by CBF-based constraints that imply (\ref{equ: rear-end safety const})-(\ref{equ:VehicleConstraints}), hence they guarantee their satisfaction, 
(b) the forward invariance property of CBFs guarantees constraint satisfaction over all future times,
(c) the new CBF-based constraints are \emph{linear} in the control, which drastically reduces the computational cost of determining them and enables real-time implementation.
In particular, based on the vehicle dynamics \eqref{VehicleDynamics}, we define $f(\bm x_i(t))=[v_i(t),0]^T, g(\bm x_i(t))=[0,1]^T$. 
Each of the constraints \eqref{equ: rear-end safety const}-\eqref{equ:VehicleConstraints} can be expressed in the form $b_j(\bm x_i(t))\geq 0, j\in 1,2, \ldots$ and the 
CBF method maps $b_j(\bm x_i(t))\geq 0$ into a new constraint which directly involves the control $u_i(t)$ and takes the form
\begin{equation}
L_{f}b_{j}(\bm x_{i}(t))+L_{g}b_{j}(\bm x_{i}(t))u_{i}(t)+\gamma_j (b_{j}(\bm %
x_{i}(t)))\geq 0 \label{equ:cbf}
\end{equation}%
where $L_{f},L_{g}$ denote the Lie derivatives of $b_{j}(\bm x_{i}(t))$ along $f$ and $g$ respectively and $\gamma_j (\cdot )$ denotes some class-$\mathcal{K}$ function (in practice, a linear class-$\mathcal{K}$ function is often used). This CBF-based constraint is a sufficient condition for $b_j(\bm x_i(t))\geq 0$ (see \cite{xiao2023safe}) and may, therefore, be conservative depending on the choice of $\gamma_j (\cdot )$.

With CBF-based constraints of the form (\ref{equ:cbf}) replacing the original ones, the new optimal control problem can be efficiently solved by discretizing time and solving a simple Quadratic Program (QP) over each time step (details can be found in \cite{xiao2023safe}) exploiting the linearity in $u$ in (\ref{equ:cbf}). One limitation of this approach is that it is ``myopic'' in the sense that the control determined at a specific time step cannot ensure that all constraints remain satisfied at future steps (e.g., they may conflict with the control constraints in (\ref{equ:VehicleConstraints})).
Given the roundabout's compact geometry which couples closely-spaced MPs, control actions propagate rapidly across different CZs and this myopic nature of control selection becomes inadequate. This motivates the 
two new elements we bring to the solution of this problem. 
First, we use a receding-horizon MPC framework to evaluate future performance not only over a single time step, but over a receding horizon $H$ of multiple lookahead steps. The choice of $H$ improves accuracy at the expense of computational cost; as further discussed in Section \ref{Sec: Simulation}, this choice can significantly affect performance, which, interestingly, is not always improving with increasing $H$.
Second, we address the problem of a CAV entering the next CZ in its path without necessarily satisfying the safety constraints required in this new CZ. We resolve this issue through the use of \emph{Control Lyapunov-Barrier Functions (CLBFs)} instead of basic CBFs, which allows us to ensure the satisfaction of these constraints within a properly selected finite time interval.

\subsubsection{MPC-based optimal control problem solution} 
Starting from current time $t$, we discretize time using time steps of equal length $T_d$. The dynamics are also discretized and we set $\bm{x}_{i,h} = \bm{x}_i(t+h\cdot T_d)$ where $h = 1, 2,\ldots$ is the $h$th time step since $t$. The decision variables $u_{i,h} = u_i (t + (h-1)\cdot T_d )$ are assumed to be constant over each time step.
Therefore, we set the control, velocity, position trajectory over next $H$ time steps as a vector respectively: $\mathbf{u}_i=[u_{i,1},\ldots u_{i,H}]$, $\mathbf{v}_i=[v_{i,1},\ldots v_{i,H}]$ and $\mathbf{x}_i=[x_{i,1},\ldots x_{i,H}]$. When adopting sequence $\bold f$, the corresponding trajectories are denoted as $\mathbf{u}_i(\bold f)$, $\bold v_i(\bold f)$ and $\bold x_i(\bold f)$.
We also denote the performance of CAV $i$ over the next $H$ time steps under a given sequence $\bold f$ as $J_i^H(\bold f)$. 
Since $J_i^H(\bold f)$ is performance over $H$ time steps, we replace travel time in objective function \eqref{eqn:energyobj} with speed multiplied by an appropriate coefficient. This optimal control problem can then be efficiently tackled by solving a simple QP over next $H$ time steps at each round.


Next, we derive the CBF constraints of the form (\ref{equ:cbf}) which replace the original constraints.

\textbf{Constraint 1} (Vehicle limitations): Let $b_1(\bm{x}_{i,h})=v_{max} - v_{i,h}$, $b_2(\bm{x}_{i,h}) = v_{i,h} - v_{min}$. Following \eqref{equ:cbf}, the corresponding CBF constraints are:
\begin{equation}\label{eqn: MPC_CBF_velocity_1}
    -u_{i,h} + \gamma_1(b_1(\bm{x}_{i,h}))\geq 0, 
    ~~\forall h \in\{1,\ldots H\}
\end{equation} 
\begin{equation}\label{eqn: MPC_CBF_velocity_2}
    u_{i,h} + \gamma_2(b_2(\bm{x}_{i,h}))\geq 0,
    ~~\forall h \in\{1,\ldots H\}
\end{equation} 

\textbf{Constraint 2} (Rear end safety constraint): Let $b_3(\bm{x}_{i,h})=x_{i_p,h}-x_{i,h}-\varphi v_{i,h} -\delta $. Similarly, the corresponding CBF constraint is:
\begin{equation}\label{eqn: MPC_CBF_rearend}
    v_{i_p,h} - v_{i,h} - \varphi u_{i,h} + \gamma_3(b_3(\bm{x}_{i,h})) \geq 0,
    ~~\forall h \in\{1,\ldots H\}
\end{equation}

\textbf{Constraint 3} (Safe merging constraint): 
The safe merging constraint \eqref{equ: Merging safe} only applies to specific time instants $t_{i_m}^k$. This poses a technical complication due to the fact that a CBF must always be in a continuously differentiable form.  We can convert (\ref{equ: Merging safe}) to a continuous time form using the technique in \cite{xiao2021decentralized} to obtain:
\begin{equation}\label{eqn:safe_merging_convert}
z_{i,i_m}(t)\geq \frac{\varphi\cdot x_{i_m}(t)}{L} v_{i}(t)+\delta,  ~~\forall t\in [t_i^{k,0}, t_{i_m}^k]   
\end{equation} 
where $t_i^{k,0}$ denotes the time CAV $i$ enters the road segment connected to $M_k$. Note that the boundary condition in \eqref{eqn:safe_merging_convert} at $t=t_{i_m}^k$ when $x_{i_m}(t) = L$ is exactly \eqref{equ: Merging safe}.
Letting $b_4(\bm{x}_{i,h})=x_{i_m,h}-x_{i,h}-\frac{\varphi}{L}x_{i_m,h}\cdot v_{i,h} -\delta$, the corresponding CBF constraint is
\begin{align}\label{eqn: MPC_CBF_merging_1}
    v_{i_m,h}&-v_{i,h}-\frac{\varphi}{L}x_{i_m,h}\cdot u_{i,h} -\frac{\varphi}{L}v_{i_m,h}\cdot v_{i,h}\nonumber\\&+\gamma_4(b_4(\bm{x}_{i,h}))\geq 0,
    ~~\forall h \in\{1,\ldots H\}
\end{align}

\textbf{Problem 2}: 
Our aim is to calculate $J_i^H(\bold f)$, $\mathbf{u}_i(\bold f)$, $\mathbf{v}_i(\bold f)$ and $\mathbf{x}_i(\bold f)$ for a single CAV $i$ given a sequence $\bold f$.
Since the travel time ($t_i^f-t_i^0$) in \eqref{eqn:energyobj} 
can no longer be directly minimized in the receding horizon optimization process, we instead maximize the speed applying a linear (similar to ($t_i^f-t_i^0$)) penalty. Since this is inversely proportional to travel time when considering the entire CAV trip, we achieve an equivalent optimization effect:
\begin{equation}
\label{eqn: MPC-CLBF with given sequence}
\begin{split}
\min_{\bm{u}_i}&~J_i^H(\bold f)=\sum_{h=1}^{H} (\frac{1}{2}u_{i,h}^2 - \lambda v_{i,h}) \\
s.t.& ~~~ \eqref{VehicleDynamics}, \eqref{eqn: MPC_CBF_velocity_1}, \eqref{eqn: MPC_CBF_velocity_2}, \eqref{eqn: MPC_CBF_rearend}, \eqref{eqn: MPC_CBF_merging_1} 
\end{split}
\end{equation}%
where $\lambda$ is a weight for balancing energy and speed.

\subsubsection{MPC-CLBF framework} 
When the constraints \eqref{equ: rear-end safety const}-\eqref{equ:VehicleConstraints} are initially satisfied, the forward invariance property of CBFs guarantees their satisfaction at all times. 
However, an unsafe initial state commonly occurs in a CZ when (\ref{equ: Merging safe}) is violated as the relative distance from the current position of CAV $i$ to the next MP is similar to the distance from CAV $i_m$ to the same MP. 
This motivates the use of \emph{Control Lyapunov-Barrier Functions (CLBFs)} where a proper choice of class-$\mathcal{K}$ function allows us to achieve finite-time convergence to a safe set if a system is initially outside this set \cite{xiao2021high}.


Although the continuous form \eqref{eqn:safe_merging_convert} is used for the safe merging constraint,
its satisfaction is only needed at the MP. Thus, if \eqref{eqn:safe_merging_convert} is not satisfied before MP, there is some time over which the CAVs $i,i_m$ can adjust their states and approach the safe set without necessarily maximally decelerating to enforce feasibility as in earlier work \cite{xu_scaling_2023}, therefore incurring expending energy and causing discomfort. 
In particular, the CLBF formulation for the safe merging constraint is:
\begin{align}\label{eqn:clbf_safe_merging_1}
    L_{f}b_4(\bm x_{i}(t))&+L_{g}b_4(\bm x_{i}(t))u_{i}(t) \nonumber\\
    &+ p\cdot b_4(\bm x_{i}(t))^q \geq 0, 
    ~~\forall t\in [t_i^{k,0}, t_{i_m}^k]  
\end{align}
where $b_4(\bm{x}_{i}(t))=x_{i_m}(t)-x_{i}(t)-\frac{\varphi}{L}x_{i_m}(t)\cdot v_{i}(t) -\delta$
and the parameters $p, q$ need to be properly selected.
Thus, (\ref{eqn:clbf_safe_merging_2}) becomes:
\begin{align}\label{eqn:clbf_safe_merging_2}
    &v_{i_m}(t)-v_i(t)-\frac{\varphi}{L}x_{i_m}(t)u_i(t) -\frac{\varphi}{L}v_{i_m}(t)v_i(t)+p  \nonumber 
    \cdot(x_{i_m}(t)\\
    &-x_i(t)-\frac{\varphi}{L}x_{i_m}(t)v_i(t)-\delta)^q \geq 0, ~~\forall t\in [t_i^{k,0}, t_{i_m}^k]
\end{align}

\textbf{$\mathbf{p,q}$ value determination.}
We define a safe merging set $C :=\{\bold x\in \mathbb{R}^2:b_4(\bold x)\geq 0\}$. If $\bold x_i(t) \in C$, we choose $p>0, q=1$, so that the last term of (\ref{eqn:clbf_safe_merging_1}) reduces to a linear class-$\mathcal{K}$ function as in the formulation of a classic CBF constraint. 
Otherwise, if $\bold x_i(t) \notin C$, we set $p>0$, $q = \frac{1}{2n+1}$, $n\in \mathbb{Z^+} $. It is proved in \cite{xiao2021high} that any control satisfying (\ref{eqn:clbf_safe_merging_2}) can force the state to converge to the safe set within time $t_m \ge t_{conv}$ where:
\begin{equation}\label{eqn:t for clbf}
    t_{conv}=\frac{{b_4(\bm x_{i}(t))}^{1-q}}{p(1-q)}
\end{equation}
Setting $\bar b(t) =\frac{\varphi}{L}x_{i_m}(t)u_{i,min} + \frac{\varphi}{L}v_{i_m}(t)v_i(t)+v_i(t)-v_{i_m}(t)$,
the following provides a sufficient condition for such finite time convergence:

\textbf{Proposition} 1: 
Given a safe set $C$ and $\bold x_i(0) \notin C$, a feasible control $u_i(t) \geq u_{i,min}$ ensures $\bold x_i(t_m) \in C$ for any $t_m \ge t_{conv}$ if $u_i(t)$ satisfies \eqref{eqn:clbf_safe_merging_2} with $p \in [\frac{{b_4(\bm x_{i}(0))}^{1-q}}{(1-q)t_m}, \frac{\bar b(0)}{{b_4(\bm x_{i}(0))}^q}]$, and $q = \frac{1}{2n+1}$, $n\in \mathbb{Z^+}$.

\begin{proof}
From \eqref{eqn:clbf_safe_merging_2}:
\begin{align}
    \label{eqn: clbf for merging: u upperbound}
    u_i(t) \leq \frac{L}{\varphi x_{im}(t)} \cdot (v_{i_m}(t)&-v_i(t)-\frac{\varphi}{L}v_{i_m}(t)v_i(t) \nonumber \\ 
    &+p\cdot {b_4(\bm x_{i}(t))}^q)
\end{align}
It is easy to verify that $\frac{\bar b(0)}{{b_4(\bm x_{i}(0))}^q}\leq \frac{\bar b(t)}{{b_4(\bm x_{i}(t))}^q}$ for $t \geq 0$. By choosing $p\leq \frac{\bar b(0)}{{b_4(\bm x_{i}(0))}^q}$ for any fixed $q$ (or $n$), we have
\begin{align}
    \label{eqn: clbf for merging: u upperbound 2}
    u_{i,min} \leq \frac{L}{\varphi x_{im}(t)} \cdot (v_{i_m}(t)&-v_i(t)-\frac{\varphi}{L}v_{i_m}(0)v_i(t) \nonumber \\ 
    &+p\cdot {b_4(\bm x_{i}(t))}^q)
\end{align}
Therefore, we can always find $u_i(t)\geq u_{i,min}$ which satisfies \eqref{eqn: clbf for merging: u upperbound} hence also \eqref{eqn:clbf_safe_merging_2}. This resolves the potential conflict between the control limit \eqref{equ:VehicleConstraints} and the CLBF constraint \eqref{eqn:clbf_safe_merging_2}.

On the other hand, with $ p \geq \frac{{b_4(\bm x_{i}(0))}^{1-q}}{(1-q)t_m}$ for any fixed $q$, we have $t_m\geq\frac{{b_4(\bm x_{i}(0))}^{1-q}}{p(1-q)}=t_{conv}$ defined in \eqref{eqn:t for clbf}, therefore convergence within $t_m$ is guaranteed. \end{proof}




In view of \textbf{Proposition} 1, we can choose proper $p, q$ values in which we set the required convergence time $t_m$ as the remaining time for CAV $i_m$ to reach the next MP (this is known since we have already calculated the trajectory of CAV $i_m$). 
Note that $n$ can be chosen as any positive integer when deciding $q$ value, and we set $n=1$ for our tests in Section \ref{Sec: Simulation}.
If we cannot find $p, q$ values satisfying \textbf{Proposition} 1, we cannot ensure that the associated control sequence is feasible and we proceed to consider the next sequence in the $F_k(t)$.

After choosing proper $p, q$ values, we can replace \eqref{eqn: MPC_CBF_merging_1} by the CLBF constraint and replace \textbf{Problem 2} as \textbf{Problem 3}: 
\begin{equation}
\label{eqn: MPC-CLBF with given sequence_2}
\begin{split}
\min_{\bm{u_i}}&~J_i^H(\bold f)=\sum_{h=1}^{H}(\frac{1}{2}u_{i,h}^2 - \lambda v_{i,h}) \\
s.t.& ~~~ \eqref{VehicleDynamics}, \eqref{eqn: MPC_CBF_velocity_1}, \eqref{eqn: MPC_CBF_velocity_2}, \eqref{eqn: MPC_CBF_rearend} \\
&v_{i_m,h}-v_{i,h}-\frac{\varphi}{L}x_{i_m,h}u_{i,h} -\frac{\varphi}{L}v_{i_m,h}v_{i,h}+p\cdot(x_{i_m,h}\\
    &-x_{i,h}-\frac{\varphi}{L}x_{i_m,h}v_{i,h}-\delta)^q \geq 0, ~~\forall h\in \{1,\ldots H\}
\end{split}
\end{equation}

The framework for determine the optimal control of a single CAV is shown in \textbf{Algorithm \ref{alg: MPC-CLBF}}.

\begin{algorithm}[!ht]
\DontPrintSemicolon
  
  \KwInput{$\bold f$, H, $i_p$, $i_m$, $\bm{u}_j(\bold f), \bm{v}_j(\bold f),\bm{x}_j(\bold f) ~~~\forall j\in \{i_p, i_m\}$}
  \KwOutput{$J_i^H(\bold f), \bm{u}_i(\bold f), \bm{v}_i(\bold f),\bm{x}_i(\bold f)$ }
  \If{Not exist $i_p$, $i_m$}
  {
    Calculate trajectories using \emph{Unconstrained Optimal Control} framework \eqref{eqn: ucp_1}-\eqref{eqn: ucp_5}
  }
  \ElseIf{Exist $i_m$}
  {
    Calculate initial safe merging condition $b_4(\bold x_i(t))$ \\
    \If{$b_4(\bold x_i(t))<0$ and $\dot b_4(\bold x_i(t))<0$ }
    {
        Sequence $\bold f$ is not feasible; \textbf{End}.
    }
    \ElseIf{$b_4(\bold x_i(t))<0$ and $\dot b_4(\bold x_i(t))
    \geq 0$}
    {
        Choose proper $p, q$ required by \textbf{Proposition} 1
    }
    \ElseIf{$b_4(\bold x_i(t))\geq 0$}
    {
        Choose proper $p>0$; $q=1$.
    }
    Solve \textbf{Problem 3}
    
  }
  \ElseIf{Exist $i_p$}{Solve \textbf{Problem 3}}
\caption{MPC-CLBF for trajectories of CAV $i$}
\label{alg: MPC-CLBF}
\end{algorithm}

The solution of this problem is obtained in a decentralized way for each CAV in a given sequence $\bold f$. This is done sequentially, so that the trajectories of previous CAVs in the sequence provide specific constraints for the following CAVs at each time step. The performance of each sequence is then evaluated as the sum of all CAV performances in this sequence.

\textbf{Determine Optimal Sequence and Motion Control}.
Once we have evaluated the performance of each feasible sequence, the optimal sequence is chosen to be the one with the best performance, i.e., $\bold f^* = \text{arg}\min_{\bold f} \sum_{i \in \bold f}J_i^H(\bold f)$.
The resulting structure of optimal sequencing over the roundabout is shown in \textbf{Algorithm \ref{alg: resequence}}.
Recall that feasible sequences are updated in event-driven fashion, triggering a re-evaluation of the optimal one and the corresponding motion control trajectories. The detailed procedures can be found in \cite{}.


\begin{algorithm}[!ht]
\DontPrintSemicolon
  
  \KwInput{H, $S(t), S_k(t), k\in \{1,2,3\}$}
  \KwOutput{$S_k^*(t),k\in\{1,2,3\}$; $\bm{u}_i^{*},\bm{v}_i^{*}, \bm{x}_i^{*}, i\in S(t)$ }
  \For{$k \in \{1,2,3\}$}    
    { 
        $J_k^{H*} = \infty$;
        $S_k^*(t) = \emptyset$;
        $\bold f^* = \emptyset$;\\
        get feasible sequence set $\bold S_k$;\\
        \For{$\bold f \in F_k(t)$}
        {
           $\hat J_k^H(\bold f)=0$;\\
           assign $i_p, i_m$ to $\forall$ CAV $i \in S_k(t)$;\\
           resequence $S_k(t)$ according to $\bold f$ and get $S_k^{\bold f}(t)$;\\
           \For{$i\in S_k^{\bold f}(t)$}
           {
               get $J_i^H(\bold f), \bm{u}_i(\bold f), \bm{v}_i(\bold f),\bm{x}_i(\bold f)$ using \textbf{Algorithm \ref{alg: MPC-CLBF}}
               
               $\hat J_k^H(\bold f) = \hat J_k^H(\bold f) + J_i^H(\bold f)$
          }
          \If{$\hat J_k^H(\bold f) < J_k^{H*}$}
          {
            $J_k^{H*} = J_k^H(c)$;
            $S_k^*(t) = S_k^{\bold f}(t)$;
            $\bold f^* = \bold f$;\\
            \For{$i\in S_k^*(t)$}
            {
                $\bm{u}_i^{*},\bm{v}_i^{*}, \bm{u}_i^{*} = \bm{u}_i(\bold f), \bm{v}_i(\bold f),\bm{x}_i(\bold f) $
            }
          }   
        }        
        }

\caption{Optimal Sequencing over a Roundabout}
\label{alg: resequence}
\end{algorithm}

\section{Simulation Results}\label{Sec: Simulation}
In this section, we use Eclipse SUMO (Simulation of Urban MObility) to model a roundabout of the form in Fig. \ref{fig:roundabout} and use its car following model
as a baseline of human-driven vehicles to compare it to the performance we obtain with CAVs under our MPC-CLBF controller. In addition, we compare the MPC-CLBF controller to the OCBF controller from our previous work \cite{xu2021decentralized}. The OCBF controller
first derives the unconstrained optimal solution as shown in Section \ref{subsubsec:Procedure for Evaluating Single CAV}. This solution is used as a reference control which is optimally tracked by minimizing quadratic control and speed deviations subject to the CBF constraints corresponding to \eqref{equ: rear-end safety const}-\eqref{equ:VehicleConstraints}. As in past work (e.g., \cite{xu2021decentralized}) this solution is done by solving a sequence of Quadratic Programs (QPs) at each discrete time step in a myopic way. This approach is computationally very efficient but its myopic nature lacks the ability to predict future CAV behavior over multiple time steps as accomplished through the MPC-CLBf approach.
Both First-In-First-Out (FIFO) and Shortest-Distance-First (SDF) sequencing policies are adopted under the OCBF controller, where SDF prioritizes a CAV with a smaller distance to the next MP and larger speed. We adopt the configuration shown in Fig. \ref{fig:roundabout} and use the same vehicle arrival patterns in all methods for consistent comparison purposes. 
The basic parameter settings are as follows: $L=60m$, $\delta=0m$, $\varphi=1.8s$, $v_{max}=30m/s, ~v_{min}=5m/s, ~u_{max}=4m/s^2, ~u_{min}=-4m/s^2$.

\textbf{Balanced traffic demand:}
 In this case, the simulated incoming traffic is generated through Poisson processes with all rates set to 396 CAVs/h, with randomly assigned exit points. The simulation time step is 0.1 s, and the total simulation time for traffic generation is 1000 s. The time horizon ($H$) for MPC indicates the number of lookahead time steps; it is set to different values for comparison purposes. 
 
 The simulation results are shown in Table \ref{table:balanced}. The total system performance for all CAVs, as well as the average performance within each CZ are recorded, combining travel time and energy consumption following \eqref{eqn:energyobj} with $\alpha=0.1$.
 The ``infeasible count'' denotes the number of times when no sequence is feasible for a merging group by solving \textbf{Problem 3}. 
 This infeasibility is always due to conflicts between CLBF constraints and control limit constraints when a new CAV enters the roundabout randomly, and no feasible $p,q$ values can be found. The ``unsafe count'' denotes the number of times when the rear-end safety constraint \eqref{equ: rear-end safety const} is violated by a CAV, which could happen due to insufficient space between a CAV and its conflicting CAV ($i_m$) at the next MP when CAV $i_m$ just crossed the MP. Note that ``unsafe'' is only in terms of the conservative speed-dependent rear-end safety constraint used by a CBF rather than a physical collision. In fact, no collision is ever observed during all of our simulation experiments. 

Table \ref{table:balanced} shows that MPC-CLBF controllers outperform the SUMO (human-driven vehicle) baseline, as well as the OCBF controller under both sequencing policies in terms of travel time, energy consumption, and safety. The total objective value under MPC-CLBF with $H=20$ shows 71.0\%, 30.2\% and 11.8\%   improvements compared to the SUMO baseline, OCBF with FIFO and SDF respectively, with the total energy consumption reduced by 94.1\%, 72.4\% and 24.5\%  respectively. 
The main reason that CAVs outperform Human-Driven Vehicles (HDVs) in terms of energy is that HDVs tend to drastically accelerate and decelerate in order to stop and wait for vehicles in another road segment to cross, or thravel through the current road segment quickly when not seeing conflict vehicles. In addition, the MPC-CLBF method outperforms OCBF since it can predict future conflicts and preempt them with smoother control actions, avoiding maneuvers with drastic control changes.

\begin{figure}[ht]
\begin{subfigure} {.48\textwidth}
  \centering
  \includegraphics[width=1\columnwidth]{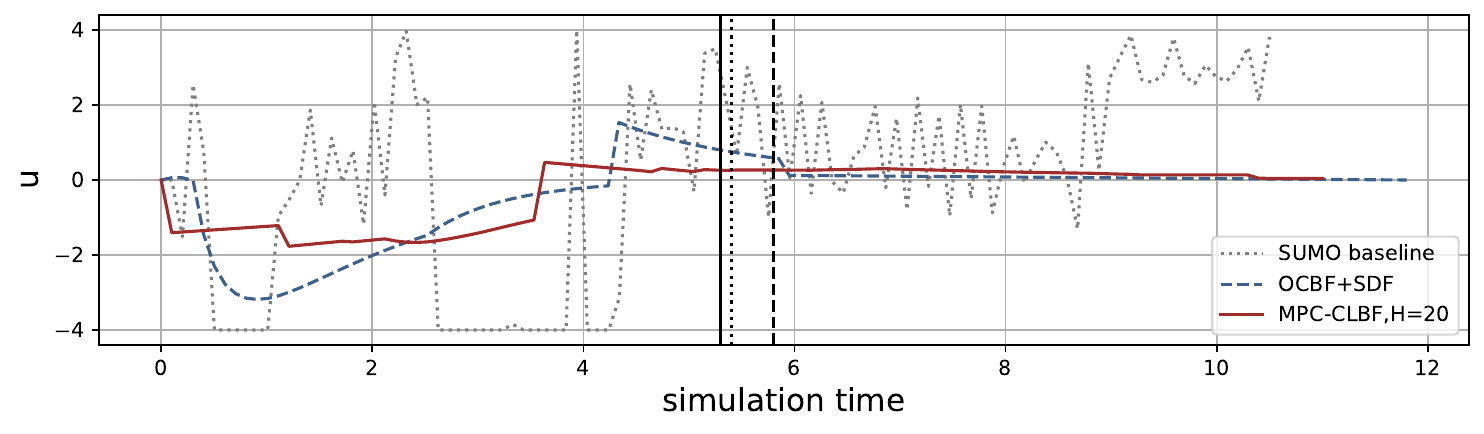} 
  \caption{control trajectories}
  \label{fig:traj_balance_cav_30_u}
\end{subfigure}
\begin{subfigure}{.48\textwidth}
  \centering
  \includegraphics[width=1\columnwidth]{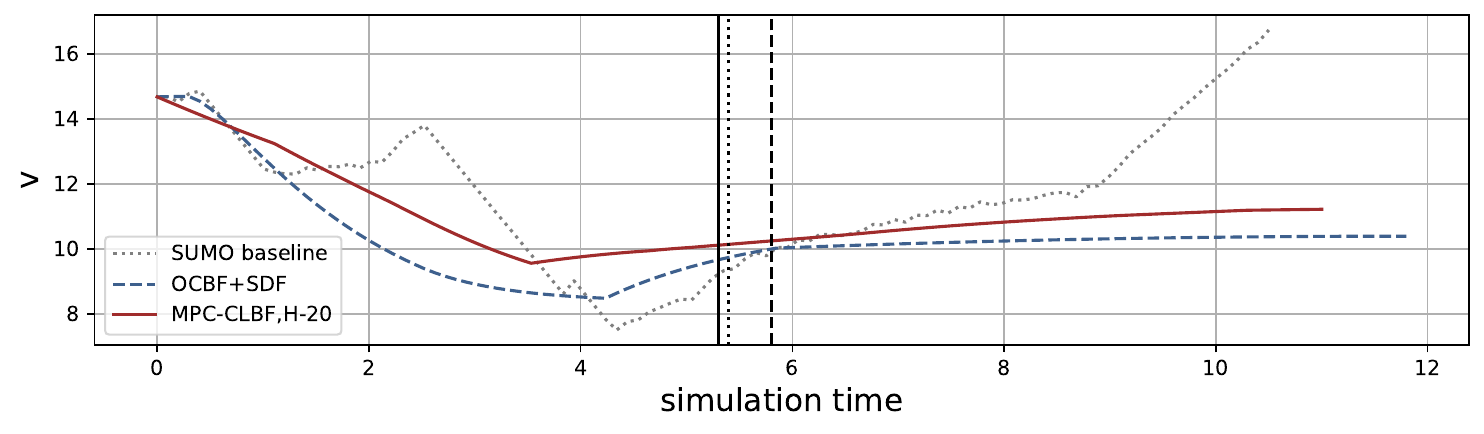}  
  \caption{velocity trajectories}
  \label{fig:traj_balance_cav_30_v}
\end{subfigure}

\caption{Comparison of trajectories from vehicle 30}
\label{fig:traj_balance_cav_30}
\end{figure}

An example of comparing trajectories for vehicle $i=30$ is shown in Fig. \ref{fig:traj_balance_cav_30}, where the red solid, blue dashed and grey dotted lines indicate the result using the MPC-CLBF method with $H=20$, OCBF with SDF, and the SUMO baseline respectively. The solid, dashed and dotted vertical lines indicate the corresponding time when vehicle 30 changes CZs.
We observe from the velocity profile that under all three methods, after entering the roundabout, vehicle 30 generally first decelerates to give way to other vehicles in the same or different road segments. It then accelerates when the conflicting vehicle exists the CZ. While the vehicles in using the SUMO car following model tend to brake and accelerate sharply as human do, the control for CAVs is much smoother. 
Compared to OCBF, speed varies less drastically with our method due to the ability of predicting future changes of other CAVs. In addition to saving energy, our method also allows faster roundabout traversal through reduced braking.


In addition, we observe that the rear-end unsafe count is reduced by more than 99\% compared to both SUMO and OCBF, which is due to the prediction ability of the MPC-CLBF control structure which can estimate future behavior and make timely control adjustments to avoid conflict while obeying the control limits in \eqref{equ:VehicleConstraints}. 

A comparison of results across different receding horizon ($H$) values reveals a trade-off between time and energy performance: While longer horizons allow CAVs to accelerate when no conflicts are foreseen, the unpredictable arrival of CAVs requires sudden harder decelerations by existing vehicles to avoid collisions, increasing the infeasibility risk. Therefore, appropriately tuning $H$ is  critical to optimize overall system performance.

\begin{table*}[]
\centering
\caption{Performance Comparison under \emph{Balanced} Traffic Demand}
\resizebox{\textwidth}{!}{%
\begin{tabular}{|l|lll|lll|lll|lll|ll|}

\hline
\multirow{2}{*}{Metric} &
  \multicolumn{3}{c|}{CZ1 average} &
  \multicolumn{3}{c|}{CZ2 average} &
  \multicolumn{3}{c|}{CZ3 average} &
  \multirow{2}{*}{total time} &
  \multirow{2}{*}{total energy} &
  \multirow{2}{*}{total obj.} &
  \multirow{2}{*}{\begin{tabular}[c]{@{}l@{}}infeasible\\  count\end{tabular}} &
  \multirow{2}{*}{\begin{tabular}[c]{@{}l@{}}unsafe\\ count\end{tabular}} \\ \cline{2-10}
 &
  time &
  energy &
  obj. &
  time &
  energy &
  obj. &
  time &
  energy &
  obj. &
   &
   &
   &
   &
   \\ \hline
SUMO Baseline &
  4.53 &
  12.96 &
  16.99 &
  4.62 &
  13.07 &
  17.18 &
  4.60 &
  12.83 &
  16.91 &
  4425.7 &
  12431.54 &
  16365.50 &
   &
  7558 \\ \hline
OCBF+FIFO &
  4.64 &
  2.44 &
  6.56 &
  4.64 &
  2.81 &
  6.94 &
  4.65 &
  2.71 &
  6.85 &
  4657.6 &
  2662.95 &
  6803.04 &
  404 &
  1544 \\
OCBF+SDF &
  4.95 &
  0.80 &
  5.20 &
  4.92 &
  1.11 &
  5.48 &
  4.96 &
  1.04 &
  5.45 &
  4958.2 &
  974.43 &
  5381.72 &
  343 &
  449 \\ \hline
MPC-CLBF, H=10 &
  5.05 &
  0.75 &
  5.23 &
  4.96 &
  0.94 &
  5.36 &
  5.08 &
  1.02 &
  5.54 &
  5033.3 &
  898.54 &
  5372.58 &
  387 &
  15 \\
MPC-CLBF, H=20 &
  4.47 &
  0.68 &
  \textbf{4.66} &
  4.50 &
  0.80 &
  \textbf{4.80} &
  4.49 &
  0.73 &
  \textbf{4.73} &
  4514.4 &
  \textbf{735.81} &
  \textbf{4748.61} &
  \textbf{256} &
  4 \\
MPC-CLBF, H=30 &
  4.25 &
  0.89 &
  4.66 &
  4.30 &
  1.09 &
  4.91 &
  4.28 &
  0.99 &
  4.80 &
  \textbf{4326.4} &
  989.59 &
  4835.28 &
  295 &
  \textbf{3} \\ \hline
\end{tabular}%
}\label{table:balanced}
\end{table*}

\begin{table*}[]
\centering
\caption{Performance Comparison under \emph{Unbalanced} Traffic Demand}
\resizebox{\textwidth}{!}{%
\begin{tabular}{|l|lll|lll|lll|lll|ll|}
\hline
\multirow{2}{*}{Metric} &
  \multicolumn{3}{c|}{CZ1 average} &
  \multicolumn{3}{c|}{CZ2 average} &
  \multicolumn{3}{c|}{CZ3 average} &
  \multirow{2}{*}{total time} &
  \multirow{2}{*}{total energy} &
  \multirow{2}{*}{total obj.} &
  \multirow{2}{*}{\begin{tabular}[c]{@{}l@{}}infeasible\\  count\end{tabular}} &
  \multirow{2}{*}{\begin{tabular}[c]{@{}l@{}}unsafe\\ count\end{tabular}} \\ \cline{2-10}
 &
  time &
  energy &
  obj. &
  time &
  energy &
  obj. &
  time &
  energy &
  obj. &
   &
   &
   &
   &
   \\ \hline
SUMO Baseline &
  4.46 &
  9.16 &
  13.12 &
  4.55 &
  14.34 &
  18.38 &
  4.61 &
  14.20 &
  18.30 &
  4629.7 &
  13004.97 &
  17120.26 &
   &
  6979 \\ \hline
OCBF+FIFO &
  4.61 &
  2.46 &
  6.56 &
  4.60 &
  1.89 &
  5.98 &
  4.67 &
  2.56 &
  6.70 &
  4815.4 &
  2389.56 &
  6669.91 &
  308 &
  1202 \\
OCBF+SDF &
  4.93 &
  0.63 &
  5.01 &
  4.87 &
  0.81 &
  5.14 &
  4.95 &
  1.09 &
  5.49 &
  5112.3 &
  886.05 &
  5430.32 &
  \textbf{263} &
  382 \\ \hline
MPC-CLBF, H=10 &
  5.10 &
  0.58 &
  5.11 &
  4.89 &
  0.79 &
  5.13 &
  5.01 &
  0.99 &
  5.45 &
  5216.0 &
  830.00 &
  5466.44 &
  304 &
  34 \\
MPC-CLBF, H=20 &
  4.55 &
  0.64 &
  4.68 &
  4.46 &
  0.66 &
  \textbf{4.63} &
  4.57 &
  0.94 &
  5.00 &
  4704.8 &
  \textbf{782.60} &
  4964.65 &
  327 &
  33 \\
MPC-CLBF, H=30 &
  4.21 &
  0.80 &
  \textbf{4.54} &
  4.22 &
  0.98 &
  4.73 &
  4.31 &
  1.15 &
  \textbf{4.98} &
  4414.80 &
  1024.91 &
  \textbf{4949.17} &
  364 &
  16 \\
MPC-CLBF, H=40 &
  4.13 &
  1.23 &
  4.90 &
  4.12 &
  1.48 &
  5.14 &
  4.21 &
  1.61 &
  5.35 &
  \textbf{4318.6} &
  1507.48 &
  5346.24 &
  383 &
  \textbf{6} \\ \hline
\end{tabular}
}\label{table:unbalanced}
\end{table*}

\begin{table*}[]
\centering
\caption{Performance Comparison under \emph{Heavy} Traffic Demand}
\resizebox{\textwidth}{!}{%
\begin{tabular}{|l|lll|lll|lll|lll|ll|}
\hline
\multirow{2}{*}{Metric} &
  \multicolumn{3}{c|}{CZ1 average} &
  \multicolumn{3}{c|}{CZ2 average} &
  \multicolumn{3}{c|}{CZ3 average} &
  \multirow{2}{*}{total time} &
  \multirow{2}{*}{total energy} &
  \multirow{2}{*}{total obj.} &
  \multirow{2}{*}{\begin{tabular}[c]{@{}l@{}}infeasible\\  count\end{tabular}} &
  \multirow{2}{*}{\begin{tabular}[c]{@{}l@{}}unsafe\\ count\end{tabular}} \\ \cline{2-10}
 &
  time &
  energy &
  obj. &
  time &
  energy &
  obj. &
  time &
  energy &
  obj. &
   &
   &
   &
   &
   \\ \hline
SUMO Baseline &
  4.96 &
  14.64 &
  19.05 &
  4.83 &
  14.96 &
  19.26 &
  4.90 &
  14.53 &
  18.89 &
  6421.4 &
  19287.16 &
  24995.07 &
   &
  13979 \\ \hline
OCBF+FIFO &
  4.84 &
  4.48 &
  8.77 &
  4.86 &
  4.55 &
  8.87 &
  4.81 &
  4.30 &
  8.58 &
  6566.9 &
  6012.82 &
  11850.06 &
  834 &
  3516 \\
OCBF+SDF &
  5.45 &
  1.56 &
  6.40 &
  5.40 &
  1.31 &
  6.11 &
  5.34 &
  1.33 &
  6.08 &
  7334.1 &
  1891.44 &
  8410.64 &
  691 &
  959 \\ \hline
MPC-CLBF, H=10 &
  5.85 &
  1.20 &
  6.40 &
  5.70 &
  1.36 &
  6.42 &
  5.76 &
  1.54 &
  6.66 &
  7813.1 &
  1841.55 &
  8786.53 &
  1360 &
  916 \\
MPC-CLBF, H=20 &
  5.18 &
  1.10 &
  5.71 &
  5.10 &
  1.32 &
  5.86 &
  5.16 &
  1.45 &
  6.03 &
  6996.0 &
  \textbf{1744.03} &
  7961.81 &
  721 &
  90 \\
MPC-CLBF, H=30 &
  4.79 &
  1.37 &
  \textbf{5.63} &
  4.80 &
  1.49 &
  \textbf{5.76} &
  4.75 &
  1.46 &
  \textbf{5.69} &
  6498.9 &
  1952.58 &
  \textbf{7729.38} &
  \textbf{609} &
  45 \\
MPC-CLBF, H=40 &
  4.64 &
  1.64 &
  5.76 &
  4.71 &
  1.86 &
  6.05 &
  4.62 &
  1.78 &
  5.89 &
  \textbf{6328.3} &
  2383.10 &
  8008.26 &
  695 &
  \textbf{36} \\ \hline
\end{tabular}%
}\label{table:busy}
\end{table*}

\textbf{Unbalanced traffic demand:} In this case, we set the traffic arrival rates to be 108 CAVs/h, 540 CAVs/h, and 540 CAVs/h for $O_1, O_2$ and $O_3$ respectively, so that the total traffic demand level for the whole roundabout is similar to the previous case. All other parameter settings are the same. The results are shown in Table \ref{table:unbalanced}. 
Unlike the previous balanced case where MPC-CLBF with $H=20$ was optimal, the total objective achieved peaks at $H=30$ using our MPC-CLBF method under these asymmetric traffic conditions, demonstrating sensitivity to the traffic demand structure. Therefore, we can appropriately tune $H$ by analyzing the historical statistics of traffic demand so as to optimize the overall system performance.
Overall, similar improvements are observed. With $H=30$, MPC-CLBF reduces the total combined travel time and energy consumption by 71.1\%, 25.8\% and 8.9\% over SUMO, OCBF with FIFO and with SDF strategies respectively. The total energy reduction indicates smoother CAV control similar to the previous case. Moreover, rear-end conflicts are dramatically reduced by over 95\% compared to other methods, exhibiting stability despite the uneven traffic.

\textbf{Heavy traffic demand:} 
It is important to assess the MPC-CLBF method's performance under heavy roundabout traffic. To that end, we increased arrival rates to 576 CAVs/h at all origins, increasing total traffic by 45.5\% over prior cases to induce congestion. All other parameters are the same. The results are shown in Table \ref{table:busy}. Owing to heavy traffic, total travel time and energy consumption increase overall compared to the previous cases. Moreover, the tight space between closely-packed CAVs traveling through the roundabout affords less room for performance improvement. 
Still, with $H=30$, MPC-CLBF reduced the total objective by 69.1\%, 34.8\% and 8.1\% compared to SUMO, OCBF-FIFO and OCBF-SDF strategies respectively. Moreover, rear-end conflicts were reduced by over 95\% compared to other methods. 
These results suggest that the MPC-CBF is consistently robust to different traffic conditions.


\textbf{Computational Complexity Analysis} 
Let the number of CZs in a roundabout be $N$ and the number of CAVs on the roundabout and entry road segments of CZ $k\in\{1,\ldots N\}$ be $N_k^0$ and $N_k^1$ respectively. Given the assumption of no vehicle overtaking, the number of feasible sequences in CZ $k$ is $\binom{N_k^0+N_k^1}{N_k^0}$. Therefore, during one round of optimal sequencing (\textbf{Algorithm \ref{alg: resequence}}), the theoretical number of times to solve \textbf{Problem 2} can be expressed as $\sum_{k=1}^N {\binom{N_k^0+N_k^1}{N_k^0} \cdot (N_k^0+N_k^1)}$, where $N_k^0$ and $N_k^1$ are bounded by the road segment length $L$. 
However, some sequences can be excluded before evaluation when no proper $p, q$ parameters for CLBF can be found. In our simulations, when $H=30$, the average number of times that \textbf{Problem 2} is solved is 4.59 and 8.54 for traffic conditions under rates of 396 CAVs/h and 576 CAVs/h respectively. The average computation time for a single solution is around $100 ms$ on an Intel Core m5 with two 1.2GHz cores using Gurobi as a numerical solver. This suggests the ability  to adopt the MPC-CLBF controller for real time implementation.

\section{Conclusion and Future Work} \label{sec: conclusion}
We presented a decentralized MPC-CLBF framework which addresses both infeasibility and myopic control issues commonly encountered when coordinating over multiple interconnected CZs. The MPC-CLBF approach is then incorporated into an Optimal Resequencing procedure, and simultaneously determines the optimal sequence and associated control jointly optimizes the future vehicle speed and the energy consumption while guaranteeing safety.
Improvements for both travel time, energy consumption and safety are shown in the simulation experiments which compare the performance of the MPC-CLBF controller to a baseline of human-driven vehicles and performance using OCBF controller. Future research is directed at sequencing and safe control over mixed-traffic scenario where human-driven vehicles are involved.

\bibliographystyle{IEEEtran}
\bibliography{ref}

\end{document}